\documentclass[a4paper]{article}

\usepackage{makecell}
\usepackage{xcolor}
\usepackage{url}
\usepackage{multirow}
\usepackage{subfigure}
\usepackage{graphicx}
\usepackage{INTERSPEECH2019}
\def\newpara{\vspace{6pt}}
\def\pretabcapt{\vspace{6pt}}
\def\prefigcapt{\vspace{2pt}}

\title{Seeing voices and hearing voices:\\ Learning discriminative embeddings using cross-modal self-supervision}
\name{Soo-Whan Chung$^{1}$, Hong-Goo Kang$^{1}$, Joon Son Chung$^{2}$}
\address{$^{1}$Department of Electrical \& Electronic Engineering, Yonsei University, South Korea\\$^{2}$ Naver Corporation, South Korea}
\email{ }

\begin{document}

\maketitle
\begin{abstract}
The goal of this work is to train discriminative cross-modal embeddings without access to manually annotated data. Recent advances in self-supervised learning have shown that effective representations can be learnt from natural cross-modal synchrony. We build on earlier work to train embeddings that are more discriminative for uni-modal downstream tasks. To this end, we propose a novel training strategy that not only optimises metrics across modalities, but also enforces intra-class feature separation within each of the modalities. The effectiveness of the method is demonstrated on two downstream tasks: lip reading using the features trained on audio-visual synchronisation, and speaker recognition using the features trained for cross-modal biometric matching. The proposed method outperforms state-of-the-art self-supervised baselines by a significant margin.
\end{abstract}
\vspace{10pt}
\vspace{10pt}


\section{Introduction}

How do we learn and understand the world around us? In contrast to the current paradigm in machine learning, humans accumulate most of their knowledge through experience rather than taught with rules and examples. Literature in perceptual studies suggest that humans rely on hearing sounds to learn about various concepts in the visual domain and vice-versa~\cite{belin2004thinking,shams2010crossmodal}. For example, most people would think that it is strange to hear a female voice from a man's face, but this is something that they usually learn through experience, not through explicit teaching.

Deep learning has led to notable success of many fields, including image recognition~\cite{simonyan2014very, liang2015recurrent}, speech recognition~\cite{graves2013speech,chan2016listen} and machine translation~\cite{bahdanau2015neural,cho2015learning}.
Fully labelled datasets already exist in these popular fields of research, but there are multiple challenges to hand-crafting datasets for every task and application -- (i) the cost of annotation can be prohibitively expensive; (ii) human annotation of data often has privacy issues, particularly those regarding biometrics; (iii) some tasks such as speaker verification can be extremely challenging even for human experts; (iv) other tasks such as emotion recognition is inherently ambiguous and cannot be clearly divided into categories. 

The Internet provides abundant resources that can be used to learn semantics and concepts, but these have not been utilised in the prevailing supervised learning framework. Recently, self-supervised learning has received a growing amount of attention with applications in many research areas~\cite{goncalves2014self,gong2017look,doersch2017multi,detone2018super}.
Most of the previous work on self-supervision are trained to reconstruct the input data~\cite{hinton2006reducing} or to predict withheld parts of the data, such as inpainting missing part of images~\cite{pathak2016context}, predicting relative position of image patches~\cite{Doersch15a} and colourising RGB images from only grey-scale images~\cite{zhang2016colorful}. 

Of closer relevance to this paper is {\em cross-modal} self-supervision, where the supervision comes from the correspondence between naturally co-occurring data streams, such as sound and images. 
Earlier works in this area have used pairwise objectives to force corresponding embeddings to be closer together and non-corresponding embeddings to be further apart using binary classification loss~\cite{Owens18,Arandjelovic18,Senocak18} or contrastive loss~\cite{Chung16a,Korbar18,tian2019contrastive}. Our recent work \cite{chung2020perfect} has demonstrated that the performance on various cross-modal tasks can be improved by re-formulating the problem as a multi-way matching task. The use of multiple negatives has also been discussed in the context of supervised metric learning~\cite{wan2018generalized,sohn2016improved} where they have also discovered advantages over pairwise objectives.

These networks are used not only for the cross-modal matching task that they have been trained on (such as audio-video synchronisation~\cite{Chung16a,chung2020perfect} or cross-modal biometrics~\cite{Nagrani18b}), but also for downstream tasks such as lip reading~\cite{Chung16a,chung2020perfect} and speaker recognition~\cite{Nagrani20d}. This shows that the self-supervised training allows representative speech and speaker features to be learnt in the individual modalities.

The goal of this work is to learn embeddings that are discriminative for the uni-modal downstream tasks, as well as for the primary cross-modal tasks that the networks have been trained on. The work naturally builds on \cite{chung2020perfect,Nagrani20d} which show that the embeddings learnt via self-supervised proxy task demonstrate promising performance on these downstream tasks. We propose a novel training function that not only penalises the distances between a video segment and non-corresponding audio segments, but also between the different video segments as well as between the different audio segments. While the previous works enforce intra-class separation and inter-class compactness {\em across} modalities, they do not penalise or optimise metrics {\em within} modalities. Therefore, the proposed cross-domain discriminative loss explicitly enforces intra-class separation {\em within} modalities in addition to the aforementioned criterion, helping the network to learn more discriminative embeddings.

The effectiveness of this method is demonstrated on three experiments each using different modalities: (1) audio-only speaker verification where the task is to verify whether or not two speech segments come from the same person, (2) audio-visual biometric matching which aims to determine whether the face image and the speech segment belongs to the same person, and (3) video-only lip reading which recognises words from silent speech. The models trained using the proposed criterion is able to produce powerful representations of both streams which are effective for all of these tasks, where we demonstrate performance that exceed state-of-the-art self-supervised baselines.


\section{Training functions}
\label{sec:model}
    
This section outlines training strategies for learning cross-modal embeddings. Existing works including AVE-Net~\cite{Arandjelovic18}, SyncNet~\cite{Chung16a} are described, followed by the cross-modal matching strategy~\cite{chung2020perfect} and our proposed variants.

\subsection{Network architecture}
Our architecture consists of two sub-networks -- one that ingests face images as input, and another that takes mel-filterbank features of speech segments. Both sub-networks are based on the VGG-M architecture~\cite{Chatfield14} which strikes a good trade-off between efficiency and performance.
The input representations and network architectures are identical for the both tasks -- audio-visual synchronisation and cross-modal biometrics. The detailed architecture is given in Table~\ref{table:architecture}.

\newpara\noindent\textbf{Audio stream.}
The inputs to the audio stream are 40-dimensional mel-filterbank coefficients in logarithm scale from sliding a hamming window of width 25ms with a stride of 10ms. The input size is $T$ frames in the time-direction, where the value of $T$ depends on the task. The network is based on the VGG-M CNN model, but the filter sizes are modified for the audio input. The layer configurations are the same as~\cite{Nagrani20d}. The network has a temporal stride of 4 in order to synchronise the sampling rate between the audio and the video streams.

\newpara\noindent\textbf{Visual stream.}
Visual stream ingests a video of cropped face, with a resolution of 224$\times$224 and a frame rate of 25 fps. The visual stream is also based on the VGG-M, but the first layer is a 3D convolution layer with a filter size of $5\times7\times7$. The feature extractor therefore has a temporal receptive field of 5 video frames and stride of 1 frame.

\begin{figure*}[ht]
    \centering
    \fbox{
    \begin{minipage}[t]{0.24\linewidth}
        \centering
        \includegraphics[width=\linewidth]{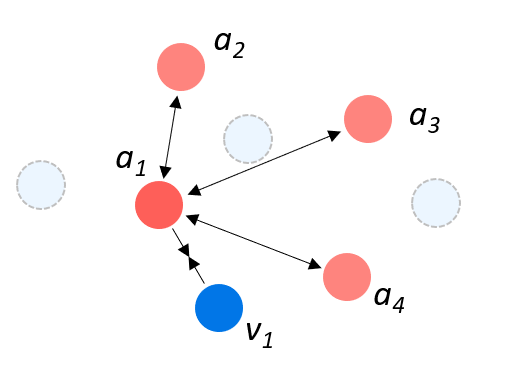}
        \centerline{(a) Audio-Audio}
    \end{minipage}
    \begin{minipage}[t]{0.24\linewidth}
        \centering
        \includegraphics[width=\linewidth]{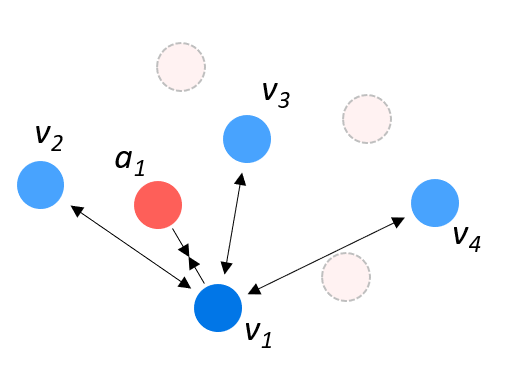}
        \centerline{(b) Video-Video}
    \end{minipage}
    \begin{minipage}[t]{0.49\linewidth}
        \centering
        \includegraphics[width=0.49\linewidth]{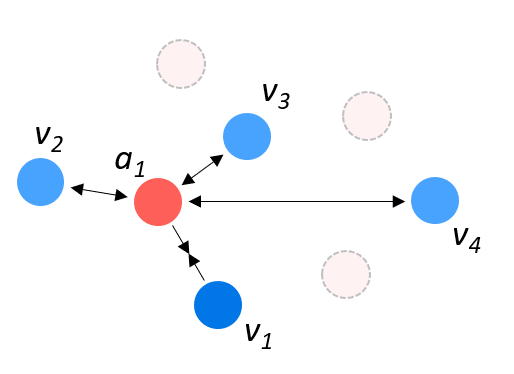}
        \includegraphics[width=0.49\linewidth]{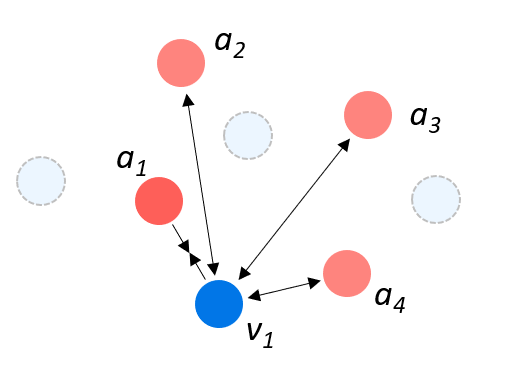}
        \centerline{(c) Audio-Video}
    \end{minipage}
    \prefigcapt
    
    }
    \caption{Cross-domain discriminative loss}
    \label{fig:metric}
\end{figure*}

\begin{table}[ht]
    \centering
    \begin{tabular}{ c | c  r | c | c  r }
        \toprule
        \multicolumn{3}{c|}{\bf (a) Audio stream} & \multicolumn{3}{c}{\bf (b) Visual stream} \\ 
        \midrule
        conv1   & [$3\times3$] & 64  & conv1 & [$5\times7\times7$] & 96\\ 
        pool1   & [$1\times1$] &     & pool1 & [$1\times3\times3$] & \\ 
        \hline
        conv2   & [$3\times3$] & 192 & conv2 & [$1\times5\times5$] & 256\\ 
        pool2   & [$3\times3$] &     & pool2 & [$1\times3\times3$] & \\
        \hline
        conv3   & [$3\times3$] & 384 & conv3 & [$1\times3\times3$] & 256\\ 
        \hline
        conv4   & [$3\times3$] & 256 & conv4 & [$1\times3\times3$] & 256\\ 
        \hline
        conv5   & [$3\times3$] & 256 & conv5 & [$1\times3\times3$] & 256\\ 
        pool5   & [$3\times3$] &     & pool5 & [$1\times3\times3$] & \\
        \hline
        conv6   & [$3\times3$] & 512 & conv6 & [$1\times6\times6$] & 512\\ 
        \hline
        fc7      &  \multicolumn{2}{c|}{512} & fc7 & \multicolumn{2}{c}{512}\\ 
        fc8     &  \multicolumn{2}{c|}{128} & fc8 & \multicolumn{2}{c}{128}\\ 
        \bottomrule
    \end{tabular}
    \pretabcapt
    \caption{Architecture of two-stream networks for audio-visual synchronisation}
    \label{table:architecture}
\end{table}

\subsection{Loss functions}

\begin{figure*}[ht]
    \centering
    \fbox{
        \includegraphics[width=\linewidth]{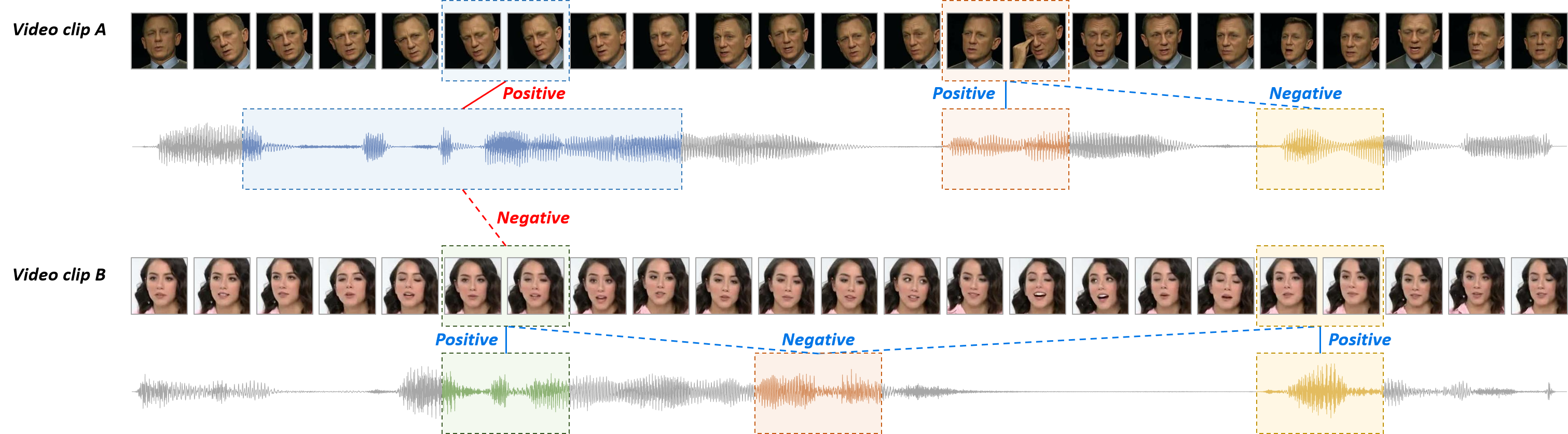}
    }
    \prefigcapt
    \caption{Sampling strategy for self-supervised training. The \textcolor{red}{red} lines represent the pairs for cross-modal biometrics, the \textcolor{blue}{blue} lines represent the pairs for lip synchronisation.}
    \label{fig:sampling}
\end{figure*}

We describe the proposed training strategies for the cross-modal embeddings, and compare it to the existing state-of-the-art methods for audio-visual correspondence. Figure~\ref{fig:sampling} gives an overview of the structure and training criteria of the existing and proposed approaches.

\newpara\noindent\textbf{Pairwise objectives.} 
SyncNet~\cite{Chung16a} and AVE-Net~\cite{Arandjelovic18} learn cross-modal correspondence using pairwise objectives. The objective is to learn representations in the joint embedding space such that the corresponding pairs are close together in the embedding space, and non-corresponding pairs are far apart. The networks are trained with contrastive loss~\cite{Chung16a} or with binary classification loss on the pairwise distances~\cite{Arandjelovic18}. For all training functions introduced in this section, it is assumed that all non-matching pairs are {\em negative} examples from different classes (e.g. different identity in the cross-modal biometrics task).

\newpara\noindent\textbf{Multi-way matching.} 
While most of the work in cross-modal learning has used pairwise objectives, \cite{chung2020perfect} proposed a network trained via a {\em multi-way matching} task. The pairwise losses only enforce that an embedding is far from one particular negative embedding, not all negatives. This algorithm adopts {\em relative} similarity of the matching pair over all non-matching pairs, leading to more stable learning. The method is related to the GE2E loss~\cite{wan2018generalized} and multi-class N-pair loss~\cite{sohn2016improved} proposed in supervised metric learning. 

For audio-to-video matching, the learning criterion compares one input feature from the audio stream to multiple features from the video stream. The task can be set up as any $N$-way feature matching task, with the objective function:
\begin{equation}
\footnotesize
L_{AV} = -\frac{1}{N} \sum_{j=1}^{N} \log
\frac{S(\mathbf{x}_A^j,\mathbf{x}_V^j)}
{\sum_{k=1}^N S(\mathbf{x}_A^j,\mathbf{x}_V^k)},
\label{eqn:AV_loss}
\end{equation}
where $\mathbf{x}_A$ and $\mathbf{x}_V$ are the audio and video embeddings respectively, and $N$ is the batch size.
For each video input, Euclidean distances between the video feature and $N$ audio features are computed, and the network is trained on the {\em inverse} of this distance. Therefore the similarity function $S$ can be written as:
\begin{equation}
\footnotesize
S({\mathbf{x}_a,\mathbf{x}_b}) = \exp(\lVert \mathbf{x}_a - \mathbf{x}_b \rVert_{2}^{-1}).
\label{eqn:euc_dist}
\end{equation}
The overall multi-way matching loss is:
\begin{equation}
\footnotesize
L_\text{MWM} = L_{AV} + L_{VA}.
\label{eqn:mwm_loss}
\end{equation}
The objective encourages the relative distance between the corresponding pairs of the audio and the videos features to be closer than the non-corresponding pairs.

\newpara\noindent\textbf{Multi-way matching with angular criterion.}
In metric learning, it has been observed that networks trained with the cosine distance metric often generalises better in unseen conditions compared to those trained with the Euclidean distance~\cite{chung2020defence,wan2018generalized}. However, the use of cosine distance combined with the softmax loss has not led to effective learning since the distances is bound to $[-1,1]$ resulting in small entropy. To resolve this, some previous works have used fixed scale parameters~\cite{wang2018cosface,deng2019arcface} to increase the dynamic range of the input to the softmax layer. In the context of metric learning, \cite{wan2018generalized} uses {\em learnable} scale parameters trained together with the network. We adopt this strategy, and replace the Euclidean metric in~\cite{chung2020perfect} with the cosine distance with learnable scale. The score function is defined as below, where $w$ and $b$ are learnable parameters.
\begin{equation}
\footnotesize
S({\mathbf{x}_a,\mathbf{x}_b}) = \exp(w\cdot \cos(\mathbf{x}_{a}, \mathbf{x}_{b})+b).
\label{eqn:cos_dist}
\end{equation}

\newpara\noindent\textbf{Cross-domain discriminative loss.}
While the existing training objectives optimise relative distances {\em across} modalities, they do not explicitly enforce discriminative power {\em within} modalities. In addition to the cross-modal matching objective in~\cite{chung2020perfect}, we introduce additional criterion that enforces the distances of different inputs from the same modality to be far apart, which enforces intra-class separation within the modality. This is particularly relevant to self-supervised representation learning, since the learnt representations are not only used for the cross-modal task, but also for uni-modal downstream tasks such as speaker verification.

In an unsupervised setting, it is not possible to obtain {\em positive} pairs from the same class. Since the features for both modalities are represented in a joint embedding space, the positive pairs are taken from the other modality, as in the cross-modal matching task. This can be formulated as:
\begin{equation}
\footnotesize
L_{AA,V} = -\frac{1}{N} \sum_{j=1}^{N} \log
\frac{S(\mathbf{x}_A^j,\mathbf{x}_V^j)}
{S(\mathbf{x}_A^j,\mathbf{x}_V^j) + \sum_{k\neq j}^N S(\mathbf{x}_A^k,\mathbf{x}_A^j)},
\label{eqn:cddl_sub_loss}
\end{equation}
where the score function of Equation~(\ref{eqn:cos_dist}) is used.
The use of positives from the other modality is comparable to mining {\em hard positives} in metric learning. 
The overall loss function is defined as below, and is also illustrated in Figure~\ref{fig:metric}.
\begin{equation}
\footnotesize
L_\text{CDDL} = L_{AV} + L_{VA} + L_{AA,V} + L_{VV,A}
\label{eqn:cddl_loss}
\end{equation}


\section{Experiments}

The models are trained two tasks and evaluated on three different tasks.

\subsection{Cross-modal biometrics}

While face and voice recognition are both widely used to verify a person's identity, the research in each field have been mostly independent due to their heterogeneous feature characteristics~\cite{reynolds2002overview,aleksic2006audio}. There has been a number of recent works that have demonstrated that a joint embedding of face and voice characteristics can be learnt for cross-modal biometric matching or verification~\cite{oh2019speech2face,kim2018learning,Nagrani18a,Nagrani18b,Nagrani20d}.

In this section, the training setup mimics that of the biometrics task of~\cite{chung2020perfect}, in which the models are trained for cross-modal biometric matching. We evaluated the trained model on two tasks: cross-modal biometric verification and audio-only speaker recognition.

\newpara\noindent\textbf{Training representations.} 
The cross-modal matching network is trained on the VoxCeleb2 dataset which consists of 5,994 different speakers with 1,092,009 clips in the training set and 118 speakers with 36,237 clips in test. The dataset is collected from YouTube clips, and the speech data comes together with face-cropped video and identity metadata. The identity labels are not used during training -- the training strategy uses audio-visual co-occurrences as the only form of supervision. We therefore assume that the audio-image pair from the same video are of the same identity, and audio-image pairs from different videos are of different identities. 

The network is trained with an auxiliary {\em content} loss on the synchronisation task, since \cite{Nagrani20d} has shown that this helps to improve performance and reduced over-fitting on the identity task by regularising features in the common encoders. 
Except for the training functions proposed in Section~\ref{sec:model}, the implementation details such as network specification, batch size and clip duration are the same as~\cite{Nagrani20d} so that the results can be compared directly.

The two stream network generates $N\times T$ identity vectors from each stream, where $N$ is the batch size and $T$ is the number of features in the time domain.
During training, the $T$ identity vectors from the audio stream are then averaged into a single vector, while a single identity vector is selected from the face stream at random. The rationale for this choice is as follows: speaker recognition requires an input of sufficient length to capture voice characteristics, but face characteristics can be obtained from a single image; we randomly select one image feature rather than average over many to prevent the model from lip reading (learning the distribution of phonemes over the time sequence).

\newpara\noindent\textbf{Evaluation on cross-modal biometric matching.}
The objective of the task is to determine whether the face image and the speech segment comes from the same person. 
Following \cite{Nagrani20d}, 300,000 pairs are audio-image pairs randomly generated from the test set of VoxCeleb2~\cite{Chung18a}, of which 10,000 are positive and the rest are negative.

\newpara\noindent\textbf{Evaluation on speaker verification.}
The task is to verify whether or not two speech segments come from the same person.
We extract identity embeddings for the data in the VoxCeleb1 \textit{test} set, which consists of 40 speakers~\cite{Nagrani17}.
The negative cosine distance between embeddings is calculated directly, then is used as the similarity score between verification pairs. 

We evaluate this on the VoxCeleb1~\cite{Nagrani17} test set. 
Cosine similarity between a pair of audio embeddings is used as the scoring function.

\begin{table}[t]
\centering 
\begin{tabular}{l r r}
\toprule
\textbf{Method} & \textbf{CBM EER} & \textbf{SV EER}  \\ 
\midrule
IL only~\cite{Nagrani20d} & 24.8\% & 23.15\% \\
IL + CL~\cite{Nagrani20d} & 23.1\%& 22.59\% \\
IL + CL + Disent. loss~\cite{Nagrani20d} & 18.9\%  & 22.09\% \\  \hline
IL + CL (Angular) & 14.9\%  & 20.25\% \\
IL + CL + CDDL (Angular)  & \textbf{13.8\%} & \textbf{17.52\%} \\
\bottomrule
\end{tabular}
\pretabcapt
\caption{Identity verification results on the VoxCeleb1 test set. Lower is better. {\bf IL} denotes the primary identity loss, {\bf CL} denotes the auxiliary content loss. For more details on the joint training, refer to~\cite{Nagrani20d}. {\bf CBM}: Cross-modal Biometric Matching; {\bf SV}: Speaker Verification.}
\label{table:results} 
\end{table}

\subsection{Visual speech recognition}

Speech and lip motion contain complementary information that is often used together by humans to interpret what is being said. In filming and broadcasting, audio-visual synchronisation is a common problem, but the solution to this problem using the speech  {\em content} has only recently received attention.

In order to synchronise the two streams, the {\em phonetic} content at a particular time in one data stream must be matched to that in the other stream. Audio-to-video synchronisation is therefore naturally a cross-modal retrieval task, where the offset is found by selecting an audio segment from a set, given a video segment. 

In this experiment, the models are trained on the cross-modal synchronisation as the proxy task and the embeddings evaluated on the video-only downstream task of visual speech recognition (lip reading).

\newpara\noindent\textbf{Training representations.} 
The self-supervised methods are trained on the {\em pre-train} set of the Lip Reading Sentences 2 (LRS2)~\cite{Chung17} dataset. The pre-train set contains 96,318 video clips containing continuous speech of various lengths. The text labels and the metadata in the LRS2 dataset are not used in any way to train the model.

Following~\cite{chung2020perfect}, audio and video segments for the matching pairs are taken at the same timestep from a facetrack; the segments for non-matching pairs are taken at different timesteps but still from the same facetrack. This is to encourage the network to learn phonetic information as opposed to biometric information.

\newpara\noindent\textbf{Evaluation on lip reading.}
The effectiveness of the trained embeddings are evaluated on the word-level classification task of the Lip Reading in the Wild (LRW) \cite{Chung16b} dataset. The dataset has a vocabulary size of 500 and over 500,000 word-level utterances, of which 25,000 are reserved for testing. The utterances are spoken by hundreds of different speakers.

For the fully supervised baselines, the VGG-M based networks are trained end-to-end on the LRW dataset.
For the pre-trained (PT) methods, the front-end feature extractors are trained with self-supervision using the methods described in Section~\ref{sec:model}, and only the back-end classifier is trained with a classification loss with supervision.

The classifiers are trained for 500-way classification and we report top-1 and top-10 accuracies on the test set. The model specifications are identical to that described in Table 2 of~\cite{chung2020perfect} to enable direct comparison of the results. 

\begin{table}[ht]
    \centering
    \begin{tabular}{ l   l  r   r }
    \toprule
        {\bf Arch.} & {\bf Method} & {\bf R@1} & {\bf R@10}\\ 
    \midrule
        MT-5     & E2E~\cite{Chung18}                  & 66.8\% & 94.6\% \\ 
        TC-5  & E2E~\cite{chung2020perfect}                  & 71.5\% & 95.9\% \\ 
        TC-5  & PT - AVE-Net~\cite{Arandjelovic18}         & 66.7\% & 94.0\% \\ 
        TC-5  & PT - SyncNet~\cite{Chung16a}         & 67.8\% & 94.3\% \\
        TC-5  & PT - Multi-way~\cite{chung2020perfect}    & 71.6\% & 95.2\% \\  \hline
        TC-5                    & PT - Multi-way (Angular)  & 73.8\% & 96.4\% \\ 
        TC-5                    & PT - CDDL (Angular)  & {\bf 75.9\%} & {\bf 97.0\%} \\ 
    \bottomrule
    \end{tabular}
    \pretabcapt
    \caption{Word accuracy of visual speech recognition using various architectures and training methods. {\bf R@K}: Recall at K or top-K accuracy.}
    \label{table:result_lr}
\end{table}  

\newpara\noindent\textbf{Results.}
The results on the lip reading experiments are reported in Table~\ref{table:result_lr}. 
When trained end-to-end with full supervision, the classification accuracy of the TC-5 model introduced in~\cite{chung2020perfect} is higher than that the existing networks of similar size such as the MT-5 model in~\cite{Chung18}.
 The results show that our proposed models outperform the baselines by a significant margin. It is notable that the self-supervised methods can outperform the end-to-end trained models even with the large-scale fully labelled dataset.


\section{Conclusions}

In this paper, we proposed a new strategy for learning discriminative cross-modal embeddings without access to manually annotated data. The method enforces not only intra-class separation of features across modalities, but also within modalities helping the model to learn more discriminative embeddings.
The model trained using the proposed objective outperforms existing methods on cross-modal biometric verification which is the task that it has been trained on,as well as on the downstream tasks such as speaker recognition and visual speech recognition. 


\section{Acknowledgements}

We would like to thank Bong-Jin Lee, Jaesung Huh, Seongkyu Mun and Hee Soo Heo
at Naver Clova for their helpful advice.

\clearpage
\raggedbottom
\bibliographystyle{IEEEtran}
\bibliography{shortstrings,mybib,vgg_local,vgg_other}
\end{document}